\documentclass[useAMS,usenatbib]{mn2e}

\usepackage{natbib}
\usepackage{amsmath,amsfonts,amssymb}
\usepackage[dvips]{graphicx}
\usepackage{longtable}
\usepackage{float}
\usepackage{caption}
\usepackage{subfig}
\usepackage{txfonts}

\title[10C Survey -- Data Analysis]
  {10C Survey of Radio Sources at 15.7~GHz: I -- Observing, mapping and source extraction
  \thanks{We request that any reference to this 
   paper cites `AMI Consortium: Franzen et al. 2010'}}
\author[AMI Consortium: Franzen et al.]
  {AMI Consortium:
  Thomas~M.~O.~Franzen,$^1$\thanks{Email: t.franzen@mrao.cam.ac.uk}
  Matthew~L.~Davies,$^1$\thanks{Email: m.davies@mrao.cam.ac.uk} 
  \newauthor  
  Elizabeth~M.~Waldram,$^1$ 
  Keith~J.~B.~Grainge,$^{1,2}$
  Michael~P.~Hobson,$^1$
  \newauthor
  Natasha~Hurley-Walker,$^1$  
  Anthony~Lasenby,$^{1,2}$
  Malak~Olamaie,$^1$
  Guy~G.~Pooley,$^1$
  \newauthor
  Carmen Rodr\'{i}guez-Gonz\'{a}lvez,$^1$  
  Richard~D.~E.~Saunders,$^{1,2}$
  Anna~M.~M.~Scaife,$^3$
  \newauthor
  Michel~P.~Schammel,$^1$
  Paul~F.~Scott,$^1$
  Timothy~W.~Shimwell,$^1$
  David~J.~Titterington,$^1$  
  \newauthor
  and Jonathan~T.~L.~Zwart$^4$\\
  $^1$Astrophysics Group, Cavendish Laboratory,
      19 J.~J.~Thomson Avenue, Cambridge CB3 0HE \\
  $^2$Kavli Institute for Cosmology Cambridge,
      Madingley Road, Cambridge, CB3 0HA \\
  $^3$Dublin Institute for Advanced Studies,
      31 Fitzwilliam Place, Dublin 2, Ireland \\
  $^4$Columbia Astrophysics Laboratory, Columbia
      University, 550 West 120th Street, New York,
      NY 10027, U.S.A.}      
\date{Accepted ????. Received ????}
\pagerange{\pageref{firstpage}--\pageref{lastpage}}
\pubyear{2010}

\begin{document}
\maketitle
\label{firstpage}

\begin{abstract}

\noindent
We have observed an area of $\approx 27$~deg$^{2}$ to an rms noise level
of $\lessapprox 0.2~\mathrm{mJy}$ at 15.7~GHz, using the Arcminute Microkelvin
Imager Large Array.  These observations constitute the most sensitive
radio-source survey of any extent ($\gtrsim 0.2$~deg$^{2}$) above 1.4~GHz.
This paper presents the techniques employed for observing, mapping and
source extraction. We have used a systematic procedure for extracting
information and producing source catalogues, from maps with varying noise
and \textit{uv}-coverage.  We have performed simulations to test our
mapping and source-extraction procedures, and developed methods for
identifying extended, overlapping and spurious sources in noisy images.
In an accompanying paper, \citet{davies2010}, the first results from the
10C survey, including the deep 15.7-GHz source count, are presented.
\end{abstract}

\nokeywords

\section{Introduction}\label{Introduction}

The 9C survey \citep{waldram2003,waldram2010} mapped 29~deg$^{2}$ of 
sky to 5.5~mJy completeness at 15~GHz, in addition to several larger
and shallower areas.  The Ryle Telescope (RT), which carried out the
survey, has subsequently been reconfigured and re-equipped to form the
Large Array (LA) of the Arcminute Microkelvin Imager \citep[AMI;][]{zwart2008}.
As part of this metamorphosis, three of the RT's eight antennas were moved to
the north of the (almost) east-west line on which they originally stood,
providing the telescope with north-south baselines.  In addition, new
front-end receivers and back-end electronics, including a new correlator,
were installed.  The result is a telescope with much larger bandwidth
and improved flux-density sensitivity, allowing us to extend our 15-GHz-band
survey work to much deeper flux-density levels.

The 9C survey was conceived to provide information regarding
the foreground radio sources that contaminated the Cosmic Microwave
Background radiation observations of the Very Small Array \citep{watson2003}.
Similarly, the 10C source survey has been designed to complement the
other AMI science programmes, which also require knowledge of contaminating
radio sources.

This paper is focussed on the techniques employed for observing, mapping
and source extraction in the 10C survey.  In an accompanying paper
\citep[hereafter Paper~II]{davies2010} the first results from the 10C
survey, including the deep 15.7-GHz source count, are presented.

\section{The Arcminute Microkelvin Imager Large Array}\label{The Arcminute Microkelvin Imager Large Array}

The AMI consists of two separate telescopes -- the LA and
Small Array (SA).  The 10C-survey data were collected using only
the LA.  Before proceeding further, the LA's vital statistics are
summarised.

The LA is an interferometer comprising eight 13-metre-diameter,
equatorially-mounted dishes, with a range of baselines of 18--110~m.
It operates at frequencies between 13.9
and 18.2~GHz with the passband divided into six channels of 0.72-GHz
bandwidth.  It has a primary beam at $\approx$~15.7~GHz of 
$\approx 5.5$~arcmin full width at half-maximum (FWHM) and a typical
resolution of $\approx 30$~arcsec (this varies depending on the
precise \textit{uv}-coverage of any observation).  The telescope
measures a single, linear polarisation (Stokes $I + Q$)
and has a flux-density sensitivity of $\approx 3$~mJy for an integration
time of one second.

Accurate pointing is important for high-frequency observations because of the relatively small telescope primary beams for such observations. The LA pointing is calibrated using five-point observations of bright point sources, which are carried out on a monthly basis. The data from five-point observations are used to construct a pointing model which provides the HA and Dec. offset of each antenna at any point in the sky. After applying these empirical corrections to the data, there is a residual rms antenna pointing error of $\approx 30$~arcsec. In practice, this is small enough compared with the primary beam size to have no significant effect on the accuracy of source flux densities measured from the raster maps.

\section{Flux-density calibration}

It is standard practice, whilst observing using the LA, to
visit a bright ($\gtrsim 200$~mJy), unresolved source close
(within a few degrees, if possible) to the target of interest for
one in every 11 minutes.  The majority of these sources are found
in the catalogue of the Jodrell-VLA Astrometic Survey
\citep[JVAS;][]{patnaik1992,browne1998,wilkinson1998}
and are used for phase and amplitude calibration of the data.

The LA is not yet sufficiently stable to rely on daily LA measurements 
of the primary calibrator sources for flux-density calibration,
and so the secondary calibrators are used for flux-density calibration.
The flux densities, in each channel, assumed for these secondary
calibrator sources are obtained from SA measurements.  The closest
(in time) SA observation of the source is selected and the 
flux densities measured in each channel are adopted as the assumed
values for calibrating the LA observation.  As many of the secondary
calibrator sources have varying flux densities, the SA observations
are made within 10 days of the relevant LA observation.  In turn,
the SA observations are calibrated by using SA observations of the
AMI primary calibration sources -- 3C286 and 3C48 -- each of
which are usually observed daily.  

The flux densities assumed for 3C286 and 3C48, in each of the AMI
frequency channels, are shown in Table~\ref{tab:fluxcal}.  The flux
densities for 3C286 were converted from total-intensity measurements
provided by R.~Perley (private communication) and made by the Very
Large Array, and are consistent with the \citet{rudy1987}
model of Mars transferred on to absolute scale, using results from the 
\textit{Wilkinson Microwave Anisotropy Probe}.  Values for 3C48
were obtained by measuring the ratios of the flux densities of 3C48 to
3C286 using SA observations carried out between 2008 January and 2010
May.

Finally, in order to account for phase errors, a correction factor of 1.082 is applied to raster flux densities. This correction factor was determined using self-calibrated maps of $\approx 50$ of the brightest sources in the survey fields -- the data were collected using pointed observations, which were carried out to check the raster maps' flux-density scale and are described in Paper II.

\begin{table}
 \caption{Assumed flux densities for sources used for primary flux-density
          calibration (channels one and two are not used routinely,
          because of interference from satellites).}
 \label{tab:fluxcal}
 \begin{tabular}{@{}c c c c }
 \hline
 Channel & $\bar{\nu}$/GHz & \multicolumn{2}{c}{$S_{I+Q}$/Jy} \\
         &                 & 3C286    & 3C48                  \\
 \hline
 3       & 13.9            & 3.74     & 1.89                  \\
 4       & 14.6            & 3.60     & 1.78                  \\
 5       & 15.3            & 3.47     & 1.68                  \\
 6       & 16.1            & 3.35     & 1.60                  \\
 7       & 16.9            & 3.24     & 1.52                  \\
 8       & 17.6            & 3.14     & 1.45                  \\
 \hline
 \end{tabular}
\end{table}

\section{Observations and rastering techniques}\label{Observations and rastering techniques}

\begin{figure}
 \includegraphics[width=0.5\textwidth,angle=270]{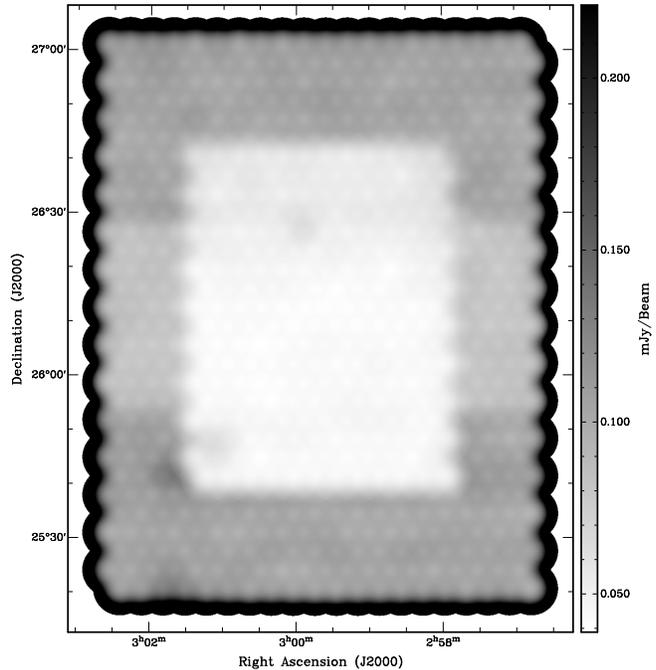}
 \caption{Noise map for one of the 10C fields.  Lighter shades indicate lower
	  noise areas.}
 \label{fig:noise_map}
\end{figure}

The 10C survey covers a total area of $\approx 27$~deg$^{2}$ divided
into 10 fields. The area within the FWHM of the LA primary
beam ($\approx 0.005$~deg$^{2}$) is a small fraction of the total
survey area.  As a result, a rastering technique, similar to that
used for the 9C survey, has been used.

Each survey field is observed using a set of telescope pointings which lie
on a 2-D hexagonally-gridded lattice, projected on to the plane of the sky.  The
fields are observed with the lattice rows running along lines of constant
declination at date.  Since the telescope is equatorially mounted, this is
the natural way to observe in order to minimise time lost to slewing between
pointings.

Maps are created for each of the
pointings and combined as described below.  No attempt has been made to
recover spatial scales larger than the primary beam by jointly deconvolving
data from separate pointings and, in this respect, the mosaicing technique
employed here is different from that of \citet{cornwell1988} and \citet{sault1996}.

The spacing between pointing centres was chosen as 4~arcmin, as this was found
(initially from simulations, but also in practice) to provide an acceptable
compromise between the desire to minimise variations in sensitivity across the
final map (achieved by increasing the number of pointing directions) and to
minimise observing time lost to telescope slewing.

The spacing between pointing centres chosen for the 10C survey (4.0~arcmin) is
smaller than for the 9C survey (5.0~arcmin) in consequence of the LA's
higher maximum observing frequency compared to the RT.  Had the same spacing
been used the variation in sensitivity across the raster maps at the
high-frequency-end of the AMI observing band would have been unacceptably
greater than that on the RT raster maps (owing to the scaling of the primary
beam with frequency).

The time spent observing each field has been apportioned so as to create
two distinct levels: the `central', with lower noise, and the `outer',
with a noise level about twice that of the central level.  This scheme
dovetails well with the SA observing programme.  Surveying the outer areas
to higher flux-density levels is also advantageous from the point of view
of improving the source-count statistics at the brighter end of the survey.
Fig.~\ref{fig:noise_map} shows the noise map for one of the 10C survey
fields.  The outer and central areas are readily apparent.

Each of the fields was divided into smaller `sub-fields', which come in two
varieties: `extended' which, taken together, cover the entirety of the field
and `core' which lie on the central area.  Since the fields have different
sizes, the number of pointings within each sub-field and the number of
sub-fields vary between fields -- the core sub-fields range between 54 and 72,
and the extended between 110 and 200, pointings.

\begin{figure}
\includegraphics[scale=0.40,angle=270]
{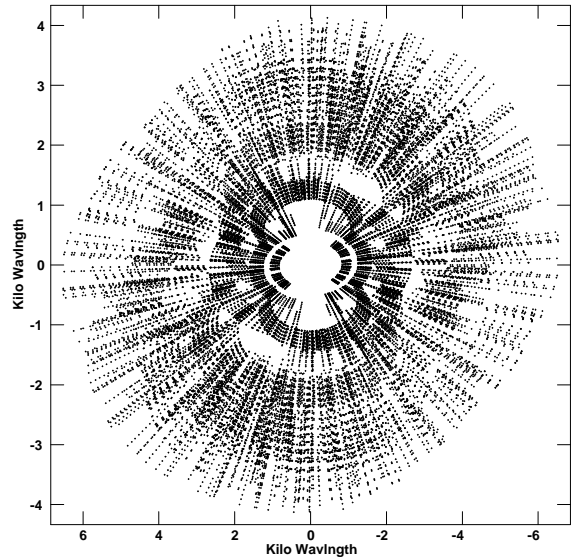}
\caption{The \textit{uv}-coverage for a typical pointing in the core area of the 
final raster; data from all individual observations of this pointing
have been combined.}
\label{fig:uv_plot}
\end{figure}

The sub-fields were each observed several times, typically for $\approx 10$\,h
at a time, until the desired noise levels were reached.  The telescope
dwell time per pointing was 30\,s for the extended and 60\,s for the core
sub-fields.  The dwell times were chosen so that in an observing run of
typical length, each pointing was visited several times, serving to
improve the \textit{uv}-coverage. Fig.~\ref{fig:uv_plot} shows
the \textit{uv}-coverage for a typical pointing in a final raster map.  Owing
to flagging of the data, for example due to poor weather conditions during
observing, the noise was sometimes found to be higher than desired over
small areas of the final map; these areas were targeted with additional,
small rasters or with individual pointings to ensure more uniform coverage
over the field.

\section{Mapping}\label{Mapping}

Raw data files from individual observing runs had various checks and
calibrations applied using in-house, data-reduction software,
\textsc{reduce}.  A  fuller explanation of the steps carried out in
\textsc{reduce}, including the weighting of data depending on the
weather conditions, can be found in \citet{davies2009}.  The data
were then output as multisource \textit{uv}-\textsc{fits} files, which were,
in turn, combined into a single multisource \textit{uv}-\textsc{fits} file per
field, again using in-house software.

Depending on the precise nature of the flagging applied,
it is possible that data belonging to individual telescope pointings
are constituted from different frequency channels in slightly varying
proportions.  As a consequence, when the data were
combined into a single \textit{uv}-\textsc{fits} file, the data weights were
adjusted, on a pointing by pointing basis, so that the continuum maps
would always have the same centre frequency: 15.7~GHz.  This causes
a small loss of sensitivity.

The \textit{uv}-\textsc{fits} data were exported into the
\textsc{aips}\footnote{\textsc{astronomical image processing system} -- www.aips.nrao.edu/} package and
individual, continuum maps created for each telescope pointing using the
\textsc{imagr} task.  Each component map is 512 by 512 pixels,
where the pixels are $5 \times 5$~arcsec$^{2}$ in size.  Natural weighting
was used to maximise signal-to-noise.

Each component map was \textsc{clean}ed using an elliptical Gaussian fitted to the
central region of the dirty beam as the restoring beam.  As a result, the
restoring beam for each component map is slightly different -- the
implications of this for source extraction are discussed in
Section~\ref{Gaussian-fitting}.  \textsc{clean}ing was stopped after the first negative
\textsc{clean} component was reached, as real negative features are not expected to
be present in the 10C maps. Negative \textsc{clean} components are in some cases necessary 
to best recover the true sky brightness distribution. However, we have chosen to adopt 
a conservative approach in order to avoid the so-called `\textsc{clean} bias' 
(\textsc{aips} cookbook; www.aoc.nrao.edu/aips/cook.html),
where real source flux is underestimated as a result of attributing \textsc{clean} 
components to noise fluctuations. The noise on each component map was then
estimated using the \textsc{imean} task, which fits a Gaussian, centred on
zero, to the distribution of pixel values.

Following a method similar to that of \cite{waldram2003}, the component maps
were combined to form a single `raster' map for each field, using the 
\textsc{flatn} task.  Only data above the 0.1 power-point of the primary
beam from each component map were used.  The mapping was also attempted
using a cut-off of 0.3, since the uncertainties in the primary beam
increase with distance from its centre.  However, this was found to
produce significant discontinuities in the map, which in turn caused spurious
source detections (see Section~\ref{Application of source extraction techniques
to real data}).  Using a cut-off of 0.1 is found to make negligible difference
to the source flux-densities and to the map noise and
avoids the problem of sharp discontinuities in the map.  For any point on
the final raster map, data from between one and seven component maps
contribute.

The map value $M_{\mathrm{r}}$ at any point on the raster map is derived
from the individual map, $m_{i}$, primary beam, $p_{i}$, and noise, $\sigma_{i}$,
values of the relevant overlapping constituent maps, at that
point.  The maps are added, correcting for the primary beam,
$\frac{m_{i}}{p_{i}}$, and weighted according to the noise,
$(\frac{p_{i}}{\sigma_{i}})^{2}$, such that

\begin{eqnarray}
M_{\mathrm{r}} = \left(\displaystyle\sum_{i=1}^{j} \frac{m_{i}p_{i}}{\sigma_{i}^{2}}\right)
\left(\displaystyle\sum_{i=1}^{j} \left(\frac{p_{i}}{\sigma_{i}}\right)^{2}\right)^{-1},
\, \mathrm{where~j~\in}~\{1,2,...,7\}. \nonumber
\end{eqnarray}

A noise map, which provides an estimate of the noise at the same point on
the raster map ($M_{n}$) is also computed, such that

\begin{eqnarray}
M_{\mathrm{n}} = \left(\displaystyle\sum_{i=1}^{j} \left(\frac{p_{i}}{\sigma_{i}}\right)^{2}\right)^{-1/2},
\, \mathrm{where~j~\in}~\{1,2,...,7\}. \nonumber
\end{eqnarray}
Pixel values on both sky and noise maps are in Janskys per beam.
The noise maps are found to provide an accurate representation of the
noise.  In particular, they are found to provide a better estimate
of the true thermal noise close to sources, than would be found
by considering the pixel values in the vicinity of sources on the
real maps.  Noise estimates computed using the latter method tend to
be biased high.

\begin{figure}
 \includegraphics[width=0.27\textwidth,angle=270]{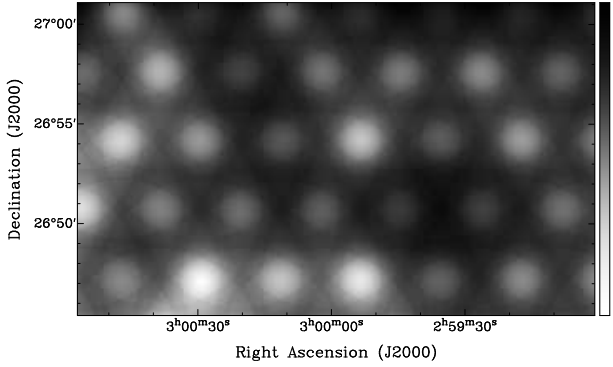}
 \caption{Detail of noise map for one of the 10C fields.  Lighter shades
	  indicate lower noise areas.  The scale varies between 96
	  and 116 $\mu$Jy.}
 \label{fig:noise_map2}
\end{figure}

Given the 4.0-arcmin spacing of the pointing centres and the 0.1 cutoff
of the primary beam, if the noise level, $\sigma_{\mathrm{c}}$, was
the same on each component map then the noise in the main area of
the raster map would vary between $\approx 0.86 \sigma_{\mathrm{c}}$ and
$\approx 0.90 \sigma_{\mathrm{c}}$.  Fig.~\ref{fig:noise_map2} is a small section of the
noise map for one of the survey fields and shows the structure
of the noise.  There is necessarily a sharp increase in the noise at the
edges of the map, as seen in Fig.~\ref{fig:noise_map}. 

\section{Source extraction}\label{Source extraction}

Sources are detected in each survey field through analysis of the
relevant raster map. For the 9C survey, candidate sources were followed up with pointed observations to allow identification of genuine sources. Observational time constraints meant that it was not possible to follow up each 10C source with a dedicated pointed observation. However, the 10C-survey noise maps provide much better estimates of the noise variation across the survey maps than were available as part of the 9C survey. Therefore, the noise maps were used to assess the reliability of each detection. Nevertheless, as described in Paper II, pointed observations were carried out towards $\approx 50$ of the brightest sources detected as part of the survey, providing a vital check of the raster maps' flux-density scale.    

In this section the steps in building a source catalogue are
described.  The methods employed are able to deal with noise levels
and synthesised beams which vary across the maps.  The source
extraction is carried out using a combination of in-house software
and \textsc{aips}.  At the end of each subsection the parameters entered
into the catalogue, at each of these steps, are provided.  Source
positions are quoted using equatorial coordinates (J2000).

\subsection{Source identification and peak measurement}\label{Source identification and peak measurement}

Source finding is carried out for each survey field making use
of the noise map, which allows sources to be identified on the
basis of their signal-to-noise ratios (SNRs).  If one wishes to
search for sources above $\gamma\sigma_{n}$, where $\sigma_{n}$
is the value of the noise map at the pixel position and $\gamma$
is a constant usually chosen to be 5.0, an initial search for
pixels greater than $0.6\gamma\sigma_{n}$ is conducted.  The value
of 0.6 was found to be sufficiently low so as to identify all peaks
that have values greater than $\gamma\sigma_{n}$ after pixel
interpolation.  Pixels selected must have values greater than or
equal to the surrounding eight pixels. Pixels at the map boundary are
not considered as potential sources.

Since the synthesised beam is fully sampled ($\sim$ 6 pixels per FWHM of the beam), 
for each candidate source
a peak value, corresponding to a position interpolated between the
grid points, is calculated. This is done by calculating the local map
values on a successively finer grid (up to 128 times finer), by 
repeated convolution with a Gaussian-graded sinc function of the form
\begin{eqnarray}
\label{eqn:map_convolution}
f(x) =
\left\{
\begin{array}{ll}
\frac{\mathrm{sin}(\pi x)}{\pi x} \mathrm{exp}(-0.275x^{2}) & \mathrm{if~} \vert x \vert \leq 3 \\
0 & \mathrm{otherwise,}
\end{array}
\right.
\end{eqnarray}
where $x$ has units of pixels \citep{rees1990}. Interpolated peak values that are
greater than or equal to $\gamma \sigma_{\mathrm{n}}$ are stored.

At this stage, the entries in the catalogue are $S_{\mathrm{pk}}$ (the
interpolated peak flux density), and $\alpha_{\mathrm{pk}}$ and
$\delta_{\mathrm{pk}}$ (the interpolated peak position).  The errors,
$\delta S_{\mathrm{pk}}$, on the source peak-flux-densities are also
provided.  They are assumed to consist of a 5 per cent calibration error added
in quadrature with the thermal noise estimate, because these are
uncorrelated.  Thus, the total error is
$\delta S_{\mathrm{pk}} = \sqrt{\sigma_{\mathrm{n}}^{2} + (0.05 S_{\mathrm{pk}})^{2}}$. 
The source names contained in the catalogue are constructed from
their respective coordinates.

\subsection{Identifying overlapping/complex sources}\label{Identifying overlapping/complex sources}

Following a method similar to that of \cite{waldram1993}, in-house software
is used to identify overlapping sources.  The integration area
of a source is taken to consist of contiguous pixels down to a lowest
contour level of 2.5$\sigma_{\mathrm{n}}$.  We begin by measuring the
integration area of the source with the lowest value of
$\sigma_{\mathrm{n}}$. The source is classified as `single' if its integration
area contains no other peaks.  Where more than one peak lies within the
integration area, they are all classified as `overlapping' and the
name of the brightest peak, as well as the number of peaks, contained
within the area are inserted in the `group' column.

The above procedure is repeated for each of the sources that
remain unclassified, in order of increasing $\sigma_{\mathrm{n}}$.
Considering the sources in this order avoids sources being classified
as single when they actually lie within the $2.5\sigma_{\mathrm{n}}$
contour of another source.

Sources can be separated by several synthesised beam widths, yet still
share the same integration area because of the presence of faint extended
emission linking them.  Overlapping sources are therefore a good
indicator of complex structure as well as point sources close together
on the sky.

\subsection{Gaussian-fitting}\label{Gaussian-fitting}

Next, the \textsc{aips} task, \textsc{jmfit} is used to fit a
2D elliptical Gaussian, by an iterative least-squares
process, to each source.  Angular sizes and integrated flux densities
can be estimated from these Gaussian fits.  For each source, the fitted
Gaussian is compared to the point-source response. This 
enables the morphology of resolved sources to be characterised. 
This also provides a
check for `over-narrow' sources, which appear significantly
smaller than the point-source response and are likely to be spurious.
Such `sources' can be caused by discontinuities in the map, for
instance at the boundary of a constituent map.  According to
\citet{condon1997} they can also occur at the positions of noise
bumps and between the peaks of overlapping or complex sources.

The comparison is complicated by the fact that the restoring beam
is slightly different for each component map, as explained in
Section~\ref{Mapping}.  As most points on the raster map result from
a combination of several constituent maps, the point-source response
varies across the raster and cannot, in general, be characterised by a
single 2-D elliptical Gaussian.  In principle, this problem
could be avoided by regrading the constituent maps with a single
beam, having dimensions that encompass all the individual restoring
beams.  However, since in practice the beam areas vary significantly
-- by up to a factor of 2.24 for the most extreme field -- this approach
has not been adopted, as this would lead to a substantial loss of 
resolution.  Rather, the point-source response belonging to the
pointing with the highest weight at the position of the source is used in the
comparison, as this proves to be a good approximation for the great
majority of sources.

Due to the large number of sources detected in the 10C survey, the 
Gaussian-fitting has been automated.  Six parameters are fitted for
each single source: the peak value, peak RA and Dec., major axis, minor
axis and position angle of the major axis.  The fitting area consists
of pixels inside a square centred on the source (peak) with a half-width
of six pixels.  Given the pixel size, this is sufficiently large to
encompass point sources (for point sources at low SNR the 2.5-$\sigma_{n}$
integration area is not sufficiently large).  An exception is made if
part of the integration area lies outside the box, in which case the
size of the box is increased such that it encompasses the integration
area, allowing correct fitting of extended sources.  The fitted
parameters are initialised as $S_{\mathrm{pk}}$, $\alpha_{\mathrm{pk}}$,
$\delta_{\mathrm{pk}}$, $b_{\mathrm{maj}}$, $b_{\mathrm{min}}$ and
$b_{\theta}$ ($b_{\mathrm{maj}}$, $b_{\mathrm{min}}$ and
$b_{\theta}$ are the parameters of the restoring beam belonging to the
pointing with the highest weight at the position of the source).  
A maximum number of 25 iterations is
allowed to reach convergence.  Even at low SNR, in real maps very few
sources did not converge within this number of iterations.  Those
sources that failed to converge were found to lie on the sidelobes
of bright sources and were often overlapping -- they were excluded from
the catalogue.

For overlapping sources of $N$ components, $N$ Gaussians are fitted simultaneously.
The peak value, peak position, major axis, minor axis and position
angle of the major axis of each component are allowed to vary. The
fitting area consists of pixels within a box encompassing
$N$ squares with half-widths of six pixels, one centred on each source
(peak). Again, an exception is made if part of the integration area lies
outside the box, in which case the size of the box is increased such that
it encompasses the integration area. The values of $S_{\mathrm{pk}}$,
$\alpha_{\mathrm{pk}}$, $\delta_{\mathrm{pk}}$, $b_{\mathrm{maj}}$,
$b_{\mathrm{min}}$ and $b_{\theta}$ for each source are used to initialise
the fit. A maximum number of 25$N$ iterations is allowed to reach
convergence.

Providing that the Gaussian-fitting achieves convergence, an integrated
flux density, $S_{\mathrm{in}}$, is obtained. This is given by
the peak height of the fitted Gaussian multiplied by the ratio of
the area of the fitted Gaussian to that of the point-source response.
For complex sources that are many times the size of the synthesised
beam, a more accurate integrated flux density can be obtained simply
by summing pixel values from the peak down to some SNR, normalising
with respect to the beam.  However, this method has
not been implemented here, since there are no believable sources in the
10C survey for which the Gaussian-fitting does not converge.  The
Gaussian-fitting also provides useful additional information.

In addition, $\alpha_{\mathrm{in}}$ and $\delta_{\mathrm{in}}$, the
fitted peak position is recorded.
Following the method of \citet{condon1997}, who has derived approximate and
semi-empirical expressions for Gaussian-fitting errors in images
with correlated noise, \textsc{jmfit} also provides an estimate of
the error, $\sigma_{\mathrm{in}}$, on $S_{\mathrm{in}}$,
due to the thermal noise.  The thermal noise, $\sigma_{\mathrm{n}}$,
is used by the task in estimating $\sigma_{\mathrm{in}}$.  Following
Section~\ref{Source identification and peak measurement}, the total
error on $S_{\mathrm{in}}$ is taken as
$\delta S_{\mathrm{in}} = \sqrt{\sigma_{\mathrm{in}}^{2} + (0.05 S_{\mathrm{in}})^{2}}$. 

An estimate of the source size is obtained by deconvolving the point-source
response from the fitted Gaussian (both represented by 2-D elliptical Gaussians).
The sky brightness distribution is represented by the resulting 2-D elliptical
Gaussian, the major and minor axes, $e_{\mathrm{maj}}$ and $e_{\mathrm{min}}$, and
position angle, $e_{\theta}$, of which are measured.  For a point source in the
limit of an infinite SNR this Gaussian will converge to
a delta function.  However, in the presence of noise the major axis will tend to
be biased high (have a value greater than 0) and the minor axis low.  In practice,
axes for which the fitted value is negative are set to zero, as negative
physical source sizes are not plausible.  Table~\ref{tab:size_interpretation}
summarises the interpretation of the major and minor axes, after deconvolution.

\begin{table}
 \caption{Interpretation of the values of the major and minor axes
          after deconvolution.}
 \label{tab:size_interpretation}
 \begin{tabular}{@{}c c l }
 \hline
 $e_{\mathrm{maj}}$  &  $e_{\mathrm{min}}$  & Interpretation                           \\
 \hline
 +ve                 &  +ve                 & Fitted Gaussian larger than point-       \\
                     &                      & source response in all dimensions.       \\
 +ve                 &  0                   & Fitted Gaussian larger than point-       \\
                     &                      & source response in some dimension.       \\
 0                   &  0                   & Fitted Gaussian equal to or smaller than \\
                     &                      & point-source response in all dimensions. \\
 \hline
 \end{tabular}
\end{table}

The results of the deconvolution can be used to classify each source as
point-like or extended.  The classification criteria are justified fully
in Section~\ref{Classifying sources as point-like or extended}.
Briefly, the source is classed as extended if $e_{\mathrm{maj}}$ is greater 
than or equal to some critical value, $e_{\mathrm{crit}}$, given by Equation~\ref{eqn:ecrit}.

The catalogue entries for the Gaussian-fitting are, therefore, $\alpha_{\mathrm{in}}$, 
$\delta_{\mathrm{in}}$, $S_{\mathrm{in}}$, $\delta S_{\mathrm{in}}$, $e_{\mathrm{crit}}$, $e_{\mathrm{maj}}$, $e_{\mathrm{min}}$ and $e_{\theta}$. A flag, $t$, is also included to denote one of
two source types: point-like (P) or extended (E). As an additional check, for each source, the integrated flux density is normalised using
the weighted sum of the areas of the point-source responses which contribute at the 
position of the source. If the resultant integrated flux density lies outside
the range $(S_{in} - \delta S_{in})$ to $(S_{in} + \delta S_{in})$ then a `*' is inserted
in the `flag' column. This indicates that the results obtained from the Gaussian-fitting
must be treated with caution, since the error from the approximation of the point-source response
is significant. In practice, this has been found to be the case for less than 1 per cent 
of sources above 5$\sigma$ detected in all survey fields.

\subsection{Example catalogue}\label{Example catalogue}

\begin{figure}
\includegraphics[scale=0.40,angle=270]
{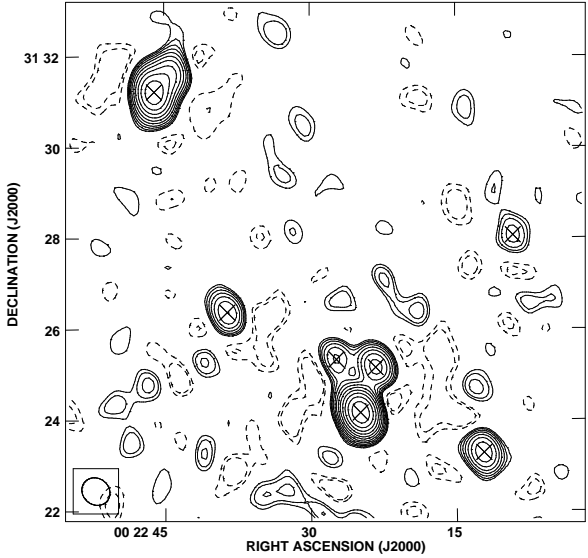}
\caption{Section of an actual map. The 1-$\sigma$ map noise is $\approx \mathrm{50~\mu Jy}$. Contour levels start at $\mathrm{\pm 100~\mu Jy}$ and increase at each level by a factor of $\sqrt{2}$.  The FWHM of the restoring beam of one of the component maps is displayed in the bottom left corner of the image. Source positions ($\alpha_{\mathrm{pk}}$,$\delta_{\mathrm{pk}}$) found with $\gamma = 5.0$ are represented as crosses.}
\label{fig:AMI001L_map_section}
\end{figure}

A contour plot of a small section of one of the survey fields, chosen since
it provides a particularly interesting example, is shown in
Fig.~\ref{fig:AMI001L_map_section}.  A subset of the catalogue entries for
the sources detected in this region are provided in Table~\ref{tab:catalogue}. Table~\ref{tab:cat_params} lists the parameters which appear in
the final source catalogues.

\begin{center}
\begin{table*}
 \caption{Abbreviated catalogue entries for sources detected in a section of an actual map (see Fig.~\ref{fig:AMI001L_map_section}). The flux-density correction accounting for phase errors, discussed in Paper II, has been applied.}
 \label{tab:catalogue}
 \begin{tabular}{@{}l c r@{} c@{} l r@{} c@{} l r r c c l}
 \hline
 Source & Group & \multicolumn{3}{|c|}{$S_{\mathrm{pk}}$~(mJy)} & \multicolumn{3}{|c|}{$S_{\mathrm{in}}$~(mJy)} &  $e_{\mathrm{crit}}$~($\arcsec$) &  $e_{\mathrm{maj}}$~($\arcsec$) &  $e_{\mathrm{min}}$~($\arcsec$) &  $e_{\theta}$~($^{\circ}$) & $t$ \\
 \hline
 
 10CJ002209+312803 & 			  &  $0.37~$  &  $\pm$  &  $~0.06$  &  $0.37~$  &  $\pm$  &  $~0.10$  & 48.9 &  9.0 &     &       & P \\
 10CJ002212+312317 & 			  &  $1.17~$  &  $\pm$  &  $~0.09$  &  $1.13~$  &  $\pm$  &  $~0.12$  & 27.5 &  8.6 &     &       & P \\
 10CJ002223+312509 & 10CJ002224+312409(3) &  $1.46~$  &  $\pm$  &  $~0.10$  &  $1.36~$  &  $\pm$  &  $~0.12$  & 25.0 &  5.3 &     &       & P \\
 10CJ002224+312409 & 10CJ002224+312409(3) &  $6.9 ~$  &  $\pm$  &  $~0.4 $  &  $7.1 ~$  &  $\pm$  &  $~0.4 $  & 25.0 &  9.1 &     &       & P \\
 10CJ002227+312520 & 10CJ002224+312409(3) &  $0.66~$  &  $\pm$  &  $~0.07$  &  $0.59~$  &  $\pm$  &  $~0.10$  & 36.5 &  4.3 &     &       & P \\
 10CJ002238+312622 & 			  &  $0.88~$  &  $\pm$  &  $~0.08$  &  $0.84~$	&  $\pm$  &  $~0.11$  & 31.0 &  7.1 &     &       & P \\
 10CJ002246+313113 & 			  &  $3.33~$  &  $\pm$  &  $~0.19$  &  $5.2 ~$  &  $\pm$  &  $~0.3 $  & 25.0 & 42.9 & 9.0 & 145.3 & E \\
  
 \hline
\end{tabular}
\end{table*}
\end{center}

\begin{table}
 \caption{The parameters which appear in the source catalogues.  The
 section in which the parameter is described is shown in brackets.}
 \label{tab:cat_params}
 \begin{tabular}{@{}l l }
 \hline
 \textit{Source}          & 10C source designation J2000 (\ref{Source identification and peak measurement}) \\
 \textit{Group}           & 10C group designation J2000 (\ref{Identifying overlapping/complex sources}) \\
 $\alpha_{\mathrm{pk}}$   & RA (peak), in h, min, s (J2000) (\ref{Source identification and peak measurement}) \\
 $\delta_{\mathrm{pk}}$   & Dec. (peak), in  \degr, \arcmin, \arcsec~(J2000) (\ref{Source identification and peak measurement}) \\
 $S_{\mathrm{pk}}$        & Peak flux density, in mJy (\ref{Source identification and peak measurement}) \\
 $\delta S_{\mathrm{pk}}$ & Error on peak flux density, in mJy (\ref{Source identification and peak measurement}) \\

 $\alpha_{\mathrm{in}}$   & RA (fitted peak), in h, min, s (\ref{Gaussian-fitting}) \\
 $\delta_{\mathrm{in}}$   & Dec. (fitted peak), in  \degr, \arcmin, \arcsec (\ref{Gaussian-fitting})  \\ 
 $S_{\mathrm{in}}$        & Integrated flux density, in mJy (\ref{Gaussian-fitting}) \\ 
 $\delta S_{\mathrm{in}}$ & Error on integrated flux density, in mJy (\ref{Gaussian-fitting}) \\

 $e_{\mathrm{crit}}$      & Critical component size, in \arcsec (\ref{Gaussian-fitting}) \\
 $e_{\mathrm{maj}}$       & Major axis after deconvolution, in \arcsec (\ref{Gaussian-fitting}) \\
 $e_{\mathrm{min}}$       & Minor axis after deconvolution, in \arcsec (\ref{Gaussian-fitting}) \\ 
 $e_{\theta}$             & Position angle after deconvolution, in \degr, \\  
                          & measured from North through East (\ref{Gaussian-fitting}) \\  
 $t$                      & Source type (P = point-like, E = extended) (\ref{Gaussian-fitting}) \\
 \textit{Flag}            & A star indicates that the approximation error for \\
                          & the point-source response is significant (\ref{Gaussian-fitting}) \\
 \hline
 \end{tabular}
\end{table}

\section{Classifying sources as point-like or extended}
\label{Classifying sources as point-like or extended}

As noted in Section~\ref{Gaussian-fitting}, criteria for classifying
sources as point-like or extended have been developed. One approach
would be to use the residuals from 3-parameter fits, where only the 
peak brightness, and peak RA and Dec. are allowed to vary, to identify
extended sources; for extended sources the fit is expected to be poor.
However, even for point sources the quality of the fit will vary with
SNR and quantifying this dependence is difficult.  An empirical approach
might be possible.  However, a method, based on theory, which makes
use of the angular size after deconvolution has been found to provide
a straightforward test applicable to a wide range of SNRs.

Maps containing point sources and noise were simulated in order to 
investigate the distribution of $e_{\mathrm{maj}}$ that would be 
expected for point sources as a function of SNR. Simulated
visibilities were generated for point sources using \textit{uv}-coverage
from real observations of a survey field.  After simulating the sources and
noise in the \textit{uv}-plane, the resultant 
multisource \textit{uv}-\textsc{fits} file was mapped as described in
Section~\ref{Mapping} and source extraction was carried out as in
Section~\ref{Source extraction}.

Point sources were simulated in the \textit{uv}-plane using a real
multisource \textit{uv}-\textsc{fits} file as a template. The
$i^{\mathrm{th}}$ visibility associated with the $j^{\mathrm{th}}$
pointing, $V_{ij}$, was set to
\begin{eqnarray}
\begin{array}{l}
V_{ij} = \displaystyle\sum_{k=1}^{N} S_{k} \exp{\phi_{ijk}} \\
\end{array}
\mathrm{,}
\end{eqnarray}
where $N$ is the total number of sources, $S_{k}$ is the
primary-beam-corrected flux density of the $k^{\mathrm{th}}$ source,
$\phi_{ijk}$ is a phase factor given by
\begin{eqnarray}
\phi_{ijk} = 2 \pi \left(u_{ij} \sin \alpha_{jk} \cos \delta_{jk} + v_{ij} \sin \delta_{jk}\right)
\mathrm{,}
\end{eqnarray}
$u_{ij}$ and $v_{ij}$ are the \textit{uv}-coordinates of $V_{ij}$,
and $\alpha_{jk}$ and $\delta_{jk}$ are the RA and Dec. separations between the
$k^{\mathrm{th}}$ source and $j^{\mathrm{th}}$ pointing centre
respectively.

A source was only added to any individual pointing if it fell
within 15~arcmin of the pointing centre.  Beyond this distance
its contribution was negligible owing to the fall-off in the
primary beam.  Gaussian noise was also simulated, such that the
noise on each of the component maps was identical, by adding random
numbers drawn from a Gaussian distribution to the real and imaginary
parts of each visibility. 

We note that point sources can alternatively be inserted into real raster maps
by adding $V_{ij}$ to real multisource \textit{uv} data. 
Such simulations were carried out in order to assess the completeness 
of the 10C survey (see Paper II).

\begin{figure}
\includegraphics[scale=0.40,angle=270]
{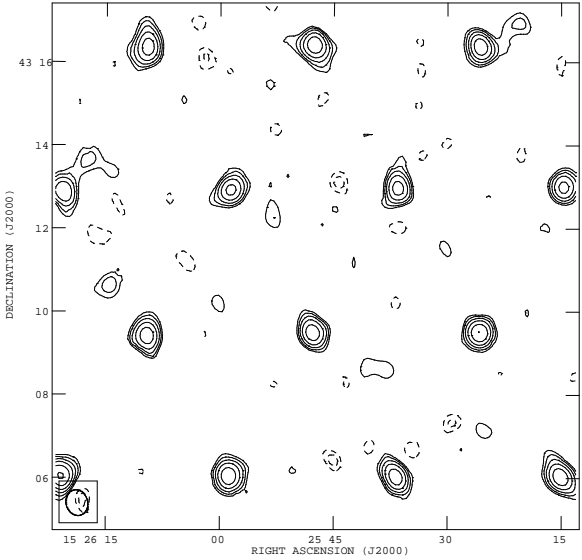}
\caption{Portion of a simulated raster map with 100-mJy point sources lying
at each pointing centre and noise about 10~mJy. Contour levels start at
$\mathrm{\pm 20~mJy}$ and increase at each level by a factor of $\sqrt{2}$.
The FWHM of the restoring beam of one of the component maps is
displayed in the bottom left corner of the image.}
\label{fig:AMI007L_PS10_m_map_section}
\end{figure}

Our source extraction techniques were applied to a map of 3.2~deg$^{2}$
containing 780, 100-mJy, simulated point sources -- one at each
pointing centre.  The rms noise on each component map was simulated to be
10~mJy per beam.  Once the component maps had been added together, as described
in Section~\ref{Mapping}, the resulting rms noise varied between $\approx$
8.6 and 9~mJy per beam over the central area of the raster map.  For illustration, 
a section of this map is shown in Fig.~\ref{fig:AMI007L_PS10_m_map_section}.

\begin{figure}
\includegraphics[scale=0.85,bb=1.20in 0.8in 5.7in 4.3in,clip=]
{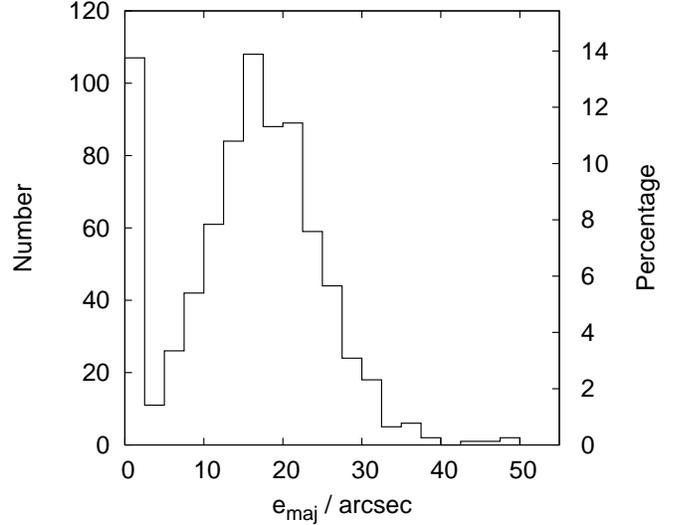}
\caption{Measured distribution of $e_{\mathrm{maj}}$ for synthetic 100-mJy point sources, with noise about 10~mJy.}
\label{fig:hist_size_PS10}
\end{figure}

Fig.~\ref{fig:hist_size_PS10} shows the distribution of
$e_{\mathrm{maj}}$ for the simulated point sources with SNRs of approximately
10.  It can be seen that it is very rare for a point source to have
$e_{\mathrm{maj}} >$ 40~arcsec at this SNR.  However, as the SNR
decreases, the distribution was found by doing runs with different source
flux densities to move further away from zero and become wider and, as a result, 40~arcsec 
would not be a suitable dividing line between point and extended sources at all SNRs.

\begin{figure}
\includegraphics[scale=0.90,bb=1.45in 0.8in 5.7in 4.3in,clip=]
{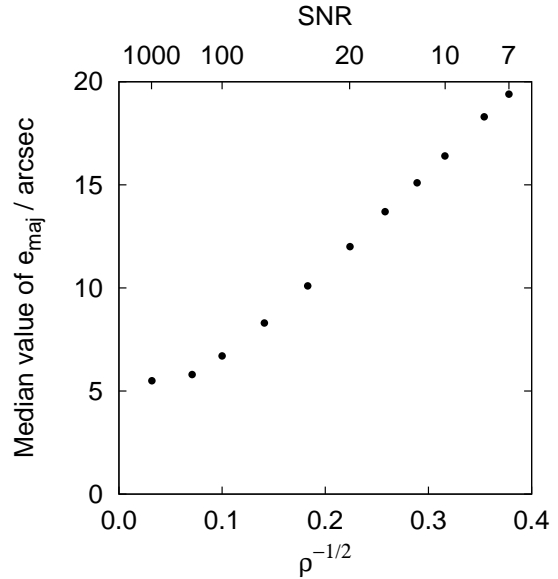}
\caption{Median value of $e_{\mathrm{maj}}$ versus $\rho^{-1/2}$ for synthetic point sources.}
\label{fig:emaj_versus_snr}
\end{figure}

The minimum component size, $e_{\mathrm{crit}}$, that can be measured is
proportional to the FWHM of the synthesised beam and inversely
proportional to the square root of the SNR \citep{fomalont2006}. 
Fig.~\ref{fig:emaj_versus_snr} shows the median value of $e_{\mathrm{maj}}$
versus $\rho^{-1/2}$, where $\rho = \frac{S_{\mathrm{pk}}}{\sigma_{\mathrm{n}}}$,
for artifical point sources with SNRs ranging between 7 and 1000.  It can be
seen that the expected relation is obtained for $\rho \lesssim 100$. The
relation starts breaking down at higher SNRs. This is probably due to the
limited level of accuracy with which we are able to simulate data, produce
maps and recover source parameters. For instance, the maps have a finite pixel
resolution which will affect the accuracy with which angular sizes can be measured.
In any case, calibration errors are the dominant source of error for 
$\mathrm{SNRs} \gtrsim 20$.

\begin{figure}
\includegraphics[scale=0.80,bb=1.15in 0.7in 5.7in 4.3in,clip=]
{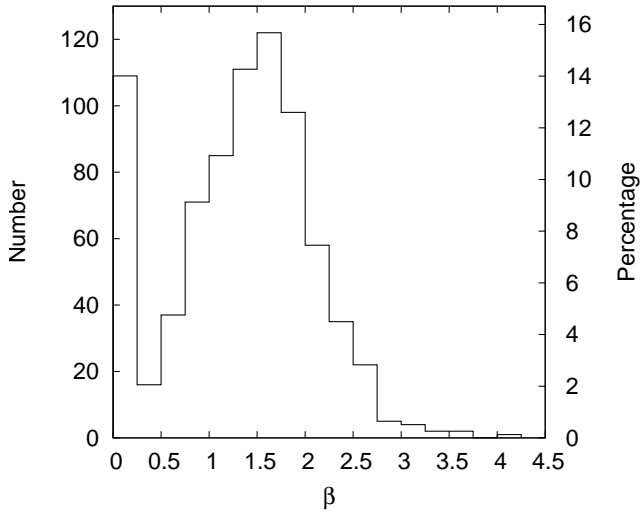}
\caption{Distribution of $e_{\mathrm{maj}}\rho^{1/2}/b_{\mathrm{maj}}$ for synthetic 100-mJy point sources, with noise about 10~mJy.}
\label{fig:hist_extent_PS10}
\end{figure}

These results indicate that
$\beta \equiv \frac{e_{\mathrm{maj}}\rho^{1/2}}{b_{\mathrm{maj}}}$
ought to be a useful indicator of source extension.
Fig.~\ref{fig:hist_extent_PS10} shows the distribution of $\beta$
for the point sources with SNRs of approximately 10.  Fewer than
1.2~per~cent of sources have $\beta > 3.0$.  In addition, the
distributions of $\beta$ for sources with SNRs ranging between 7 and
100 were investigated.  In each case, less than 1.5~per~cent of
sources were found to have $\beta > 3.0$.  As noted above, at high
SNRs, calibration errors will become the main source of uncertainty.  As a result,
sources with $e_{\mathrm{maj}} < 25$~arcsec are never considered to be extended,
no matter how high the SNR.  This value corresponds approximately to the critical
value at a SNR of 20, above which calibration errors begin to dominate.  On the
basis of our findings, a source is classified as extended if
$e_{\mathrm{maj}} \geq e_{\mathrm{crit}}$,
where
\begin{eqnarray}
\label{eqn:ecrit}
e_{\mathrm{crit}} =
\left\{
\begin{array}{ll}
3.0 b_{\mathrm{maj}} \rho^{-1/2} & \mathrm{if~} 3.0 b_{\mathrm{maj}} \rho^{-1/2} > 25.0~\mathrm{arcsec,} \\
25.0~\mathrm{arcsec} & \mathrm{otherwise.}
\end{array}
\right.
\end{eqnarray}

\begin{figure}
\includegraphics[scale=0.80,bb=1.25in 0.7in 5.7in 4.3in,clip=]
{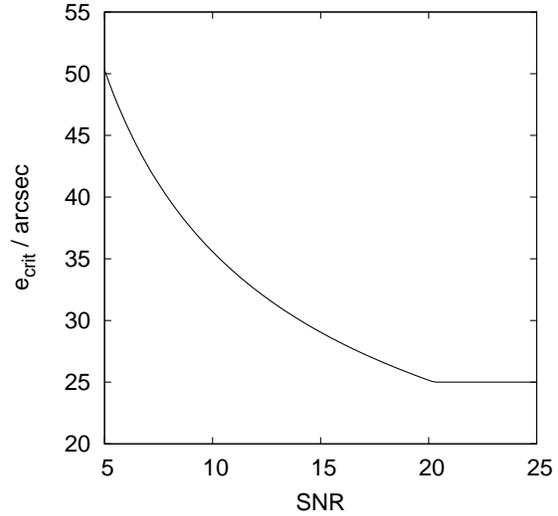}
\caption{The critical component size versus SNR for $b_{\mathrm{maj}} = 37.5$~arcsec.}
\label{fig:ecrit_versus_snr}
\end{figure}

Clearly, there is no sharp dividing line between point and extended sources,
and no single method for making the distinction. However, the above procedure
for estimating source sizes is simple and robust, and is able to deal with
sources with a wide range of SNRs. The expected percentage of point sources which 
are misclassified as extended in this scheme is $\lessapprox 1$ per cent.  
The minimum angular size that can be measured is estimated as a function of both the 
telescope resolution and the SNR. The median value of $b_{\mathrm{maj}}$ for each survey 
field varies between 32.5 and 42.6~arcsec.; it is generally lower for a field lying at a
higher declination. Fig.~\ref{fig:ecrit_versus_snr} shows how 
$e_{\mathrm{crit}}$ varies with SNR for $b_{\mathrm{maj}} = 37.5$~arcsec. 
 
The method for distinguishing between point and extended sources can be used
to establish how best to measure the total 
flux density of a source. Clearly, $S_{\mathrm{in}}$ should be used to estimate
the total flux density of an extended source. Both $S_{\mathrm{pk}}$ and
$S_{\mathrm{in}}$ can be used to describe a point source. However, at low SNR, 
the results obtained using $S_{\mathrm{in}}$ will be significantly
less accurate given the fact that the source is not assumed to be point-like when
measuring $S_{\mathrm{in}}$. In summary, $S_{\mathrm{pk}}$ should be used to describe
all sources except those that are classified as extended, for which $S_{\mathrm{in}}$
should be used.

\section{Application of source extraction techniques to real data}
\label{Application of source extraction techniques to real data}

\begin{figure}
\includegraphics[scale=0.85,bb=1.30in 0.8in 5.7in 4.3in,clip=]
{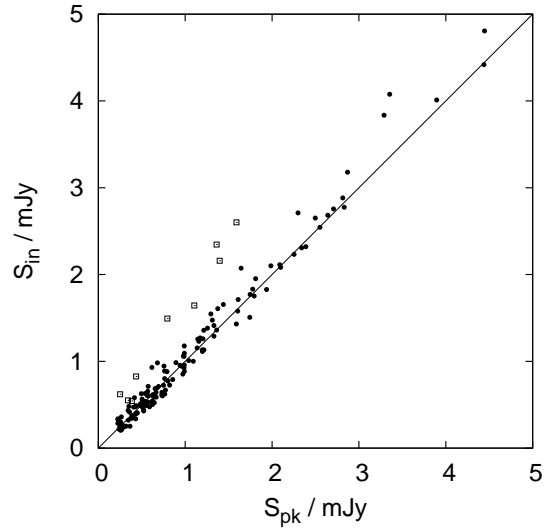}
\caption{Comparison of integrated flux densities with peak flux densities for sources detected in one of our survey fields, with a line indicating equal flux-density values. The brightest sources have been omitted from the plot. Sources classified as point-like and extended are represented as filled circles and open squares respectively.}
\label{fig:AMI002L_comp_flux}
\end{figure}

\begin{figure}
\includegraphics[scale=0.40,angle=270]
{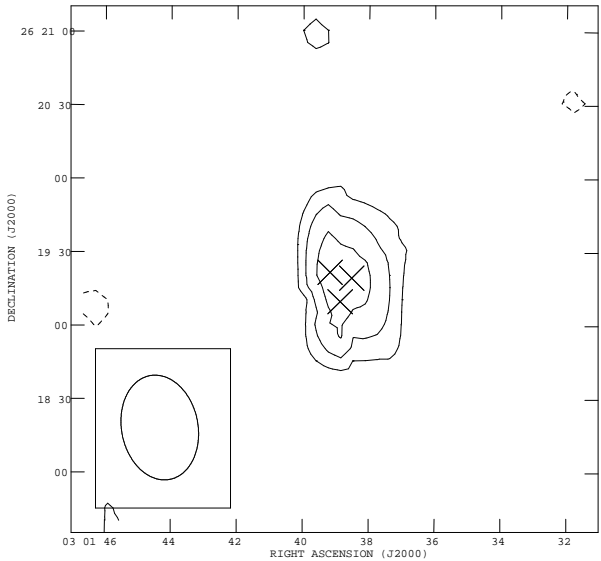}
\caption{A real source with two spurious detections when mapping was carried out using a 0.3 cut-off of the primary beam.  The 1-$\sigma$ map noise is $\mathrm{105~\mu Jy}$. Contour levels start at $\mathrm{\pm 210~\mu Jy}$ and increase at each level by a factor of $\sqrt{2}$.  Recovered peak source-positions are indicated with crosses.  The FWHM of the restoring beam of one of the component maps is displayed in the bottom left corner of the image.}
\label{fig:spurious_sources}
\end{figure}

To illustrate the source extraction procedures outlined above they
were applied to one of the survey fields, covering an area of about
2.5~$\mathrm{deg}^{2}$; the results are presented here.  The noise
levels in the core and extended field areas are typically 50 and
$100~\mu \mathrm{Jy}$ respectively.  In total, 158 sources were
detected above $5 \sigma$, of which 10 are classed as overlapping.

The Gaussian-fitting converged successfully for all sources within
25 iterations.  Fig.~\ref{fig:AMI002L_comp_flux} shows how
$S_{\mathrm{in}}$ compares with $S_{\mathrm{pk}}$ for all
sources, 10 of which are classified as extended.

No over-narrow sources were detected.  However, when
using a primary beam cut-off of 0.3 to map the data several sources
were found to have integrated flux-densities significantly smaller
than their peak flux-densities.  This is indicative of a spurious
source.  These detections were found to be separated from other
sources (spurious or real) by a distance much smaller than a
synthesised beam FWHM.  This is implausible and resulted from the
edge of a component map passing through a genuine source -- this
resulted in a sharp discontinuity in pixel values
(Fig.~\ref{fig:spurious_sources}).

\section{Conclusions}

In order to investigate the high-frequency-radio sky we have observed
10 survey fields, covering $\approx 27$~deg$^{2}$ to an rms noise level
of $\lessapprox 0.2~\mathrm{mJy}$, using the AMI LA at 16~GHz.  In an
accompanying paper, \citet{davies2010}, we present the first results
from the survey, including the deep 15.7-GHz source count.  Here, we have
concentrated on developing techniques for producing and analysing the
survey raster maps.  In particular, we have:
\begin{enumerate}

\item[(1)] developed systematic, automated methods for identifying and
characterising sources in maps with varying noise levels and sythesised
beams.

\item[(2)] proposed a straightforward and robust method for
distinguishing between point and extended sources over a wide range
of SNRs.  Our method has been tested using maps including simulated
sources and noise, and has been shown to be successful in identifying
extended emission.

\item[(3)] applied our techniques to real sky maps and demonstrated
that our automated techniques are useful for identifying and
chracterising complex structure
(see, for example, Fig.~\ref{fig:AMI001L_map_section}).

\end{enumerate}

\section*{Acknowledgments}\label{acknowledgements}

We are grateful to the staff of the Mullard Radio Astronomy Observatory for the maintenance and operation of the AMI. We thank the referee, Elaine Sadler, for helpful comments. We are also grateful to the University of Cambridge and PPARC/STFC for funding and supporting the AMI. MLD, TMOF, MO, CRG, TWS and MPS are grateful for support from PPARC/STFC studentships.

\setlength{\labelwidth}{0pt}

\label{lastpage}

\begin{thebibliography}{}

\bibitem[\protect\citeauthoryear{Browne et al.}{1998}]{browne1998}
Browne~I.~W.~A., Wilkinson~P.~N., Patnaik~A.~R., Wrobel~J.~M.,
1998, MNRAS, 293, 257

\bibitem[\protect\citeauthoryear{Cornwell}{1988}]{cornwell1988}
Cornwell~T.~J., 1988, A\&A, 202, 316

\bibitem[\protect\citeauthoryear{Condon}{1997}]{condon1997}
Condon~J.~J., 1997, PASP, 109, 166

\bibitem[\protect\citeauthoryear{AMI Consortium: Davies et al.}{2009}]{davies2009}
AMI Consortium: Davies~M.~L. et al., 2009, MNRAS, 400, 984

\bibitem[\protect\citeauthoryear{AMI Consortium: Davies et al.}{2010}]{davies2010}
AMI Consortium: Davies~M.~L et al., 2010, in preparation

\bibitem[\protect\citeauthoryear{Fomalont}{2006}]{fomalont2006}
Fomalont~E., 2006, Lecture at Tenth Summer Synthesis Imaging Workshop, University of New Mexico, USA

\bibitem[\protect\citeauthoryear{Patnaik et al.}{1992}]{patnaik1992}
Patnaik~A.~R., Browne~I.~W.~A., Wilkinson~P.~N., Wrobel~J.~M., 1992,
MNRAS, 254, 655

\bibitem[\protect\citeauthoryear{Rees}{1990}]{rees1990}
Rees~N., 1990, MNRAS, 244, 233

\bibitem[\protect\citeauthoryear{Rudy et al.}{1987}]{rudy1987}
Rudy~D.~J., Muhleman~D.~O., Berge~G.~L., Jakosky~B.~M.,
Christensen~P.~R., 1987, Icarus, 71, 159

\bibitem[\protect\citeauthoryear{Sault et al.}{1996}]{sault1996}
Sault~R.~J., Staveley-Smith~L., Brouw~W.~N., 1996, A\&AS, 120, 375

\bibitem[\protect\citeauthoryear{Waldram \& Riley}{1993}]{waldram1993}
Waldram~E.~M., Riley~J.~M., 1993, MNRAS, 265, 853

\bibitem[\protect\citeauthoryear{Waldram et al.}{2003}]{waldram2003}
Waldram~E.~M., Pooley~G.~G., Grainge~K.~J.~B., Jones~M.~E.,
Saunders~R.~D.~E., Scott~P.~F., Taylor~A.~C., 2003, MNRAS, 342, 915

\bibitem[\protect\citeauthoryear{Waldram et al.}{2010}]{waldram2010}
Waldram~E.~M., Pooley~G.~G., Davies~M.~L., Grainge~K.~J.~B.,
Scott~P.~F., 2010, 404, 1005

\bibitem[\protect\citeauthoryear{Watson et al.}{2003}]{watson2003}
Watson R.A. et al., 2003, MNRAS, 341, 1057

\bibitem[\protect\citeauthoryear{Wilkinson et al.}{1998}]{wilkinson1998}
Wilkinson~P.~N., Browne~I.~W.~A., Patnaik~A.~R., Wrobel~J.~M., Sorathia~B,
1998, MNRAS,300, 790

\bibitem[\protect\citeauthoryear{AMI Consortium: Zwart et al.}{2008}]{zwart2008}
AMI Consortium: Zwart~J.~T.~L. et al., 2008, MNRAS, 391, 1545

\end{thebibliography}
\end{document}